\documentclass[runningheads]{llncs}
\usepackage[utf8]{inputenc}
\usepackage[T1]{fontenc}
\usepackage{orcidlink}
\usepackage{graphicx}

\usepackage{makecell, multirow} 
\begin{document}
\title{Feasibility of Multifrequency Neural Network-Based Wave Inversion in Ultrasound Time-Harmonic Elastography}

%
\titlerunning{Multifrequency Neural Network-Based Wave Inversion in Ultrasound THE}
%

\author{
Héloïse Bustin\inst{1,2}\orcidlink{0009-0003-7315-3673} \and
Tom Meyer\inst{2}\orcidlink{0000-0002-2171-6791} \and
Paul Spiesecke\inst{2}\orcidlink{0000-0002-4016-9615} \and
Jakob Jordan\inst{2}\orcidlink{2222-3333-4444-5555} \and
Anja Hennemuth\inst{1,2,3,4}\orcidlink{0000-0002-0737-7375}  \and
Ingolf Sack\inst{2}\orcidlink{0000-0003-2460-1444}
}

\authorrunning{H. Bustin et al.}
%

\institute{
Deutsches Herzzentrum der Charité, Institute of Computer-assisted Cardiovascular Medicine, Augustenburger Platz 1, 13353 Berlin, Germany \\
\email{heloise.bustin@charite.de} \and
Charité – Universitätsmedizin Berlin, corporate member of Freie Universität Berlin and Humboldt Universität zu Berlin, Charitéplatz 1, 10117 Berlin, Germany \and
Fraunhofer MEVIS, Berlin, Germany \and
DZHK (German Centre for Cardiovascular Research), Berlin, Germany
}

\maketitle 
\begin{abstract}

Ultrasound time-harmonic elastography (THE) is a ~non-invasive imaging technique that measures soft-tissue stiffness. Relying on the same principle of time-harmonic shear wave excitation as established magnetic resonance elastography (MRE), THE offers full field-of-view stiffness mapping at short acquisition times and sub-second temporal resolution across organs. However, reconstruction of shear wave speed maps from acquired tissue wavefields remains challenging due to speckle noise and sparse spatial coverage of the lateral component. 

Deep learning methods have shown promising results for image reconstruction in transient-based ultrasound elastography modalities but remain unexplored in THE. In MRE, MF-ElastoNet, a 2D multifrequency neural network-based wave reconstruction method trained on synthetic physics-based wave data, has shown high generalizability across dynamic ranges and resolutions.

We investigated the potential of a multifrequency neural network-based wave inversion for THE, by performing a first feasibility analysis. To this end, we adapted the processing pipeline of MF-ElastoNet from MRE to THE. We tested our approach on THE acquired phantom data from the \textit{algorithms for mapping elastic modulus} Challenge, as well as on \textit{in vivo} liver THE in 4 healthy volunteers and 14 patients with focal liver lesions, including benign focal nodular hyperplasia, and malignant hepatocellular carcinoma and adenocarcinoma. MF-ElastoNet was successfully adapted to THE with \textit{in vivo} liver testing. Compared with k-MDEV, the state-of-the-art wave inversion method for THE, Bland-Altman analysis showed a mean bias of -0.02\,m/s (95\% limits of agreement: $-0.36$ to $0.33$\,m/s).

\end{abstract}

\section{Introduction}

Ultrasound elastography is a non-invasive imaging technique that measures soft tissue stiffness \cite{IngolfSchaefterBook2018}. Already standard of care in the staging of liver fibrosis \cite{Tzsch_tzsch_2016}, ultrasound elastography is widely evaluated and validated in tumor characterization \cite{Sigrist_2017}, including focal liver lesions \cite{Spiesecke_2026} and thyroid nodules \cite{Sigrist_2017}, as well as in chronic diseases, including chronic kidney disease \cite{Sigrist_2017} and heart failure \cite{Meyer_2025}.

Different types of ultrasound elastography techniques include strain imaging and shear wave imaging based on acoustic radiation force impulses (ARFI) or on continuous vibrations as in time-harmonic elastography (THE). Relatively novel THE is of particular interest as it allows a wide field-of-view (FOV), high penetration depth (up to 15 cm), millisecond temporal resolution and inherent comparability to MRE. It captures external vibrations at multiple frequencies through conventional B-Mode ultrasound at framerates of 80 to 100 Hz \cite{Klemmer_Chand_a_2024}. From the displacement wavefields obtained, a wave inversion reconstruction is performed, obtaining a shear wave speed (SWS) mapping over the full FOV as a proxy for tissue stiffness. Noise and unknown boundary conditions render this reconstruction ill-posed and highly challenging. Furthermore, state-of-the-art reconstruction methods, including the wavenumber-based multifrequency elasto-visco inversion (k-MDEV) \cite{Tzsch_tzsch_2016} must be tuned to the acquisition settings used for different ultrasound transducers and in different organs.

Deep learning methods in ultrasound elastography have shown promising results in solving the inverse problem in strain imaging \cite{Wu_2018,Gao_2019,Delaunay_2020,K_Z_Tehrani_2020,Yin_2024} and in ARFI-based elastography \cite{Kafaei_Zad_Tehrani_2024,Zhang_2025}. Furthermore, in MRE, where external time-harmonic excitation is widely used, several promising approaches for deep learning-based wave inversion have been developed \cite{Murphy_2017,Scott_2020,Ma_2025,Bustin_2025b,Bustin_2025a,Bustin_2027}. However, to the best of our knowledge, there have as of yet been no neural network-based investigations for wave inversion in ultrasound THE.

In this work, we performed the first feasibility analysis of a multifrequency neural network-based wave inversion method for ultrasound THE. To this end, we adapted MF-ElastoNet \cite{Bustin_2025a,Bustin_2027}, an uncertainty-aware, multifrequency neural network-based method developed for MRE, to ultrasound THE. We tested the model on phantom datasets from the \textit{algorithms for mapping elastic modulus} (aMEM) Challenge \cite{Harris_2024}, analyzing the ability of the model to resolve homogeneous regions of interest (ROIs) as well as inclusions of different stiffnesses. We additionally tested our approach on \textit{in vivo} liver THE in both healthy volunteers and patients with focal liver lesions, including benign (focal nodular hyperplasia) and malignant (hepatocellular carcinoma and adenocarcinoma) cases. In these datasets, we compared our method with k-MDEV \cite{Tzsch_tzsch_2016}, the state-of-the-art wave inversion method for THE.

\section{Methods}

\subsection{Data processing}

MF-ElastoNet is a 2D multifrequency neural network-based wave inversion approach originally developed for MRE \cite{Bustin_2025a,Bustin_2027}. It maps a multifrequency displacement wavefield to the associated multifrequency SWS and a measure of its uncertainty in a combined processing pipeline. Following an initial preprocessing of the wavefields, the neural network performs the reconstruction in a patch-based approach. Sets of multicomponent and multifrequency 4$\times$4 image patches are sampled from the wavefield and mapped to multifrequency shear wavelength values in pixels. These are then multiplied by the associated resolution and frequencies to obtain the SWS. The model was trained using a flexible framework of between 1 and 10 input frequencies \cite{Bustin_2025a,Bustin_2027}. As a first feasibility analysis, we used MF-ElastoNet as a pre-trained model, modifying only preprocessing steps to reflect the nature of THE data.

The wavefields were obtained from the ultrasound radiofrequency data using Kasai’s displacement estimation algorithm, followed by a temporal Fourier transform to isolate the induced vibration frequency components \cite{Tzsch_tzsch_2016,Meyer_2025}, and scan conversion to resample to 0.5 mm isotropic Cartesian voxel size \cite{Meyer_2022_Valsalva}. A modified version of the wavefield preprocessing used by MF-ElastoNet for MRE \cite{Bustin_2025a} was then performed. Gaussian smoothing with a standard deviation of 2 mm was applied. The smoothing strength was lowered in comparison to the original preprocessing \cite{Bustin_2025b} due to the higher resolution of THE. The Laplacian unwrapping and Fourier transform with bin selection, which were part of the original preprocessing for MRE \cite{Bustin_2025a}, were not included as they are not necessary for THE.

MF-ElastoNet was trained on 4$\times$4 image patches for wavelengths between 2.5 and 120 pixels \cite{Bustin_2025a,Bustin_2027}. However, at the frequencies and resolution at which THE was performed, the wavelengths present within a 4$\times$4 patch would be expected to be up to 3 times longer than the training range. As a first feasibility study, we chose not to modify the original model, but to extract larger patches of 12$\times$12 and to downsample these to 4$\times$4. This downsampling mechanism allowed us to obtain data within the original training range, without losing on the resolution of the final image, which would be the case when downsampling the full image.

Finally, MF-ElastoNet provided a spatially resolved measure of the uncertainty of its predictions, allowing the assessment of their trustworthiness. We reported the uncertainty through the coefficient of variation (CV), ensuring its independence with respect to the predicted SWS.

\begin{figure*}
\centering
\includegraphics[width=\textwidth]{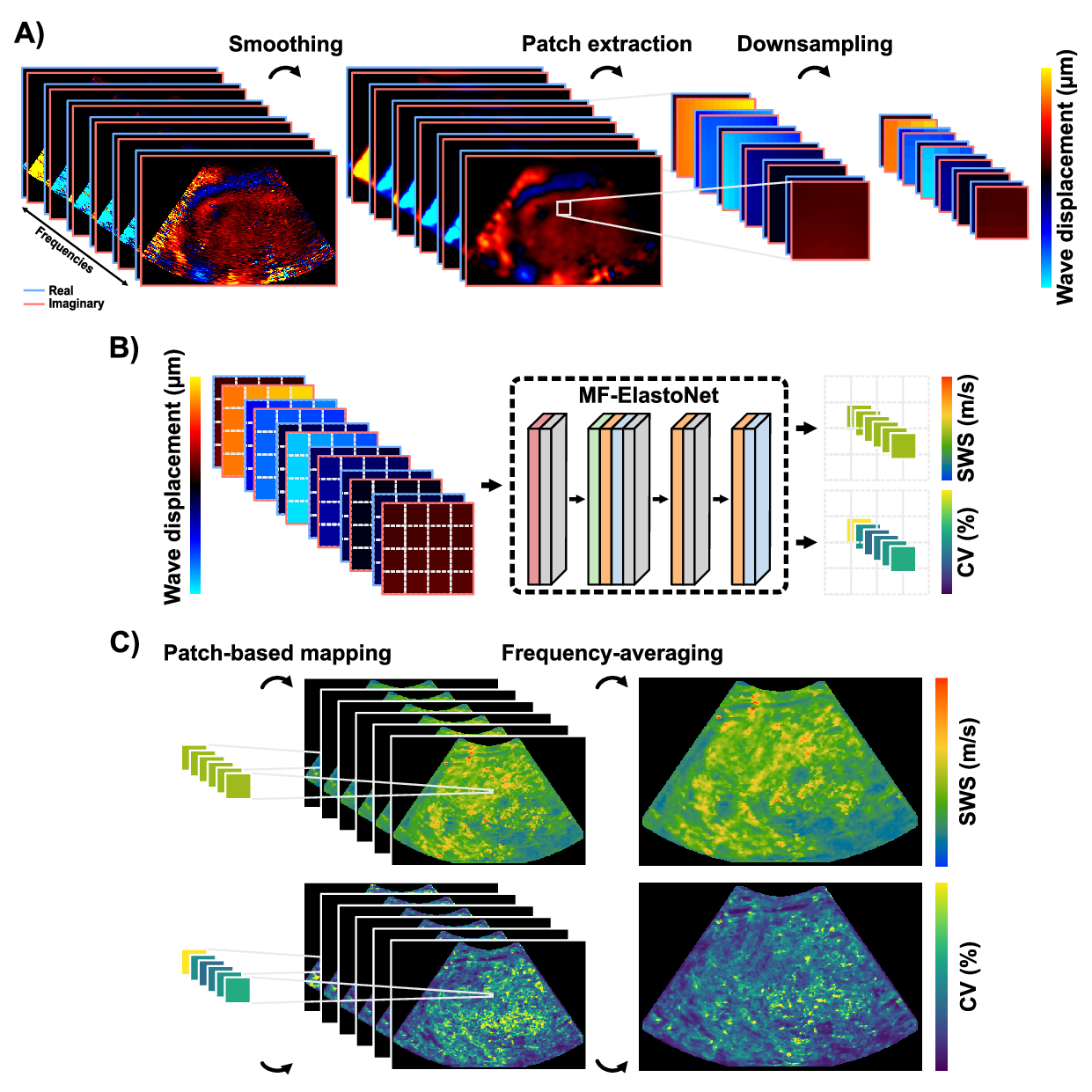}
\caption{Wave inversion processing pipeline: A) Multifrequency complex-valued wavefields were smoothed. 12$\times$12 patches were then extracted from the images and downsampled to 4$\times$4. B) Multifrequency 4$\times$4 complex-valued patches were processed by MF-ElastoNet, obtaining multifrequency SWS (shear wave speed) and quantifying the associated prediction uncertainty via the CV (coefficient of variation). C) Processed SWS values and associated uncertainty were mapped to frequency-resolved SWS and CV maps. These were averaged to obtain the final SWS and CV maps.} \label{fig:method}
\end{figure*}

Figure \ref{fig:method} presents the wave inversion processing pipeline. The adapted processing pipeline of MF-ElastoNet for THE has been made publicly available on the open-access elastography platform \url{https://bioqic-apps.charite.de/} \cite{Meyer_2022}.

\subsection{Testing datasets}

\subsubsection{aMEM Challenge dataset}

The aMEM Challenge dataset consists of measurements of three CIRS 049 phantoms. Phantom I contains a homogeneous region, phantom II a 16.68 mm diameter spherical stiff inclusion at a 15 mm depth, and phantom III a 20 mm diameter spherical soft inclusion at a 35 mm depth. These data were acquired using a C5-2v convex array transducer (Verasonics, Inc., Kirkland, WA) and a Verasonics 64 system (Verasonics, Inc., Kirkland, WA) with radiofrequency sampling at 14.28 MHz and at harmonic multifrequency vibration of 100, 120, 140, 160, 180, and 200 Hz. These data are fully described in \cite{Harris_2024} and are publicly available on the open-access elastography platform \url{https://bioqic-apps.charite.de/} \cite{Meyer_2022}.

As there was no ground truth available for these phantoms, we used SWS image quality metrics as proposed in the challenge. In the homogeneous phantom I, spatial variation of SWS was assessed through the standard deviation of the SWS. Phantoms II and III were assessed based on the contrast-to-noise ratio (CNR), edge sharpness (through the edge width), and edge accuracy. The evaluation script, including the regions of interest (ROIs) and the metrics implementation provided by the challenge, was used for this evaluation \cite{Harris_2024}. A comparison with k-MDEV as submitted to the challenge was performed.

\subsubsection{\textit{In vivo} liver THE}

We evaluated our approach on \textit{in vivo} liver THE in 4 healthy volunteers (4 males; age: $33.7 \pm 12.0$ years) \cite{Meyer_2022_Valsalva} and 14 patients with focal liver lesions, including benign focal nodular hyperplasia (FNH; 4 females; age: $50.5 \pm 15.7$ years) and malignant hepatocellular carcinoma (HCC; 1 female, 4 males; age: $73.1 \pm 7.7$ years) and adenocarcinoma (1 female, 4 males; age: $57.7 \pm 20.0$ years) \cite{Spiesecke_2026}. In both studies, liver THE was performed using the commercially available THE system (GAMPTmbH, Merseburg, Germany) equipped with a 3.3 MHz convex array transducer. The radiofrequency data were acquired at a framerate of 80 Hz with a penetration depth of 13 cm over 1 s using continuous harmonic vibrations at six superimposed frequencies (27, 33, 39, 44, 50, and 56 Hz). Controlled aliasing was performed to recover the 44, 50, and 56 Hz frequencies due to the Nyquist limit of 40 Hz imposed by the framerate as in \cite{Tzsch_tzsch_2016,Tzsch_tzsch_2018}. ROIs were manually segmented by a trained radiologist from the B-mode images. A comparison with the SWS maps obtained by k-MDEV \cite{Spiesecke_2026} was performed, including a Bland-Altman analysis.

\section{Results}

\subsection{aMEM Challenge dataset}

\begin{figure*}
\centering
\includegraphics[width=0.8\textwidth]{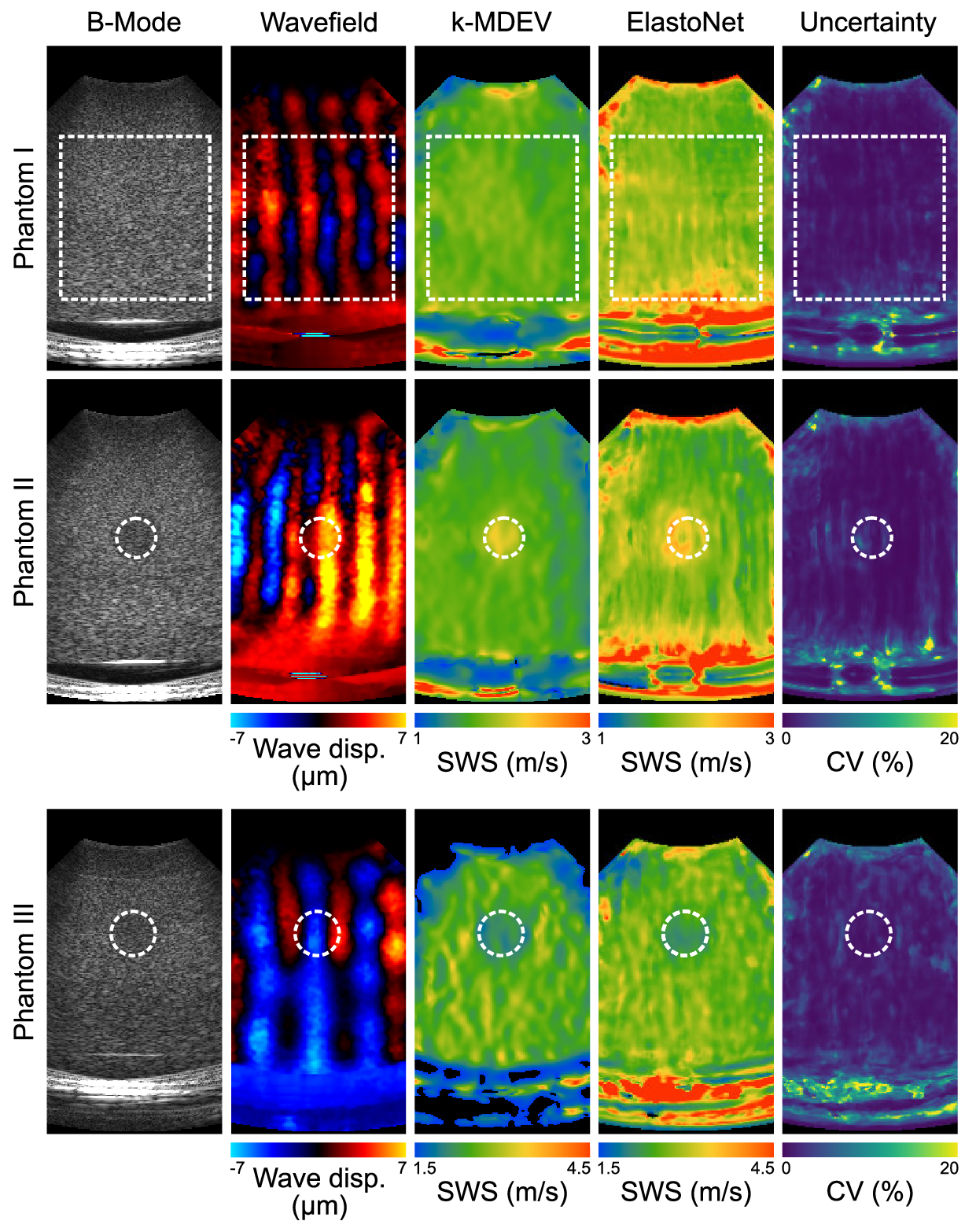}
\caption{Wavefields, SWS (shear wave speed), and CV (coefficient of variation) maps predicted by MF-ElastoNet for phantoms I, II, and III as well as their respective ROIs (regions of interest) as a white dashed line. } \label{fig:amem_challenge_images}
\end{figure*}

The wavefield, SWS, and uncertainty maps for phantoms I, II, and III are displayed in Figure \ref{fig:amem_challenge_images}, and we report the challenge metrics in Table \ref{tab:amem_challenge_metrics}. For MF-ElastoNet, phantom I appeared homogeneous within the identified ROI, except in the far field, where artifacts due to low signal-to-noise ratio (SNR) induced some inhomogeneities. This was well reflected in the uncertainty map, as the CV also increased in those areas. The map produced by k-MDEV had a smoother appearance than that produced by MF-ElastoNet, consistent with its lower standard deviation. For MF-ElastoNet the circular inclusion was clearly identifiable in phantoms II and III. The soft inclusion of phantom III showed better contrast and edge accuracy than the stiff inclusion of phantom II, whereas phantom II had better edge sharpness than phantom III. Across both phantoms, the greater detail captured in the SWS maps produced by MF-ElastoNet relative to k-MDEV was reflected in the higher edge sharpness and edge accuracy in phantom II as well as in the higher CNR and edge accuracy in phantom III.

\begin{table}
\caption{aMEM Challenge metrics for k-MDEV and MF-ElastoNet: For phantom I, the spatial variation of the SWS (shear wave speed) is reported through the standard deviation (std.). For phantoms II and III, the CNR (contrast-to-noise ratio), edge sharpness through the edge width, and the edge accuracy are reported.}
\label{tab:amem_challenge_metrics}
\centering
\begin{tabular}{|c|c|c|c|}
\hline
Phantom & Metric & k-MDEV & MF-ElastoNet \\
\hline
Phantom I & SWS std. (m/s) & \textbf{0.083} & 0.178 \\
\hline
\multirowcell{3}{Phantom II} & CNR & \textbf{2.87} & 1.57 \\
& Edge width (mm) & 8.16 & \textbf{3.66} \\
& Edge accuracy (mm) & -2.37 & \textbf{1.25} \\
\hline
\multirowcell{3}{Phantom III} & CNR & 2.09 & \textbf{2.69} \\
& Edge width (mm) & \textbf{2.93} & 6.55 \\
& Edge accuracy (mm) & 1.69 & \textbf{0.64} \\
\hline
\end{tabular}
\end{table}

\subsection{\textit{In vivo} liver THE}

\begin{figure*}
\centering
\includegraphics[width=\textwidth]{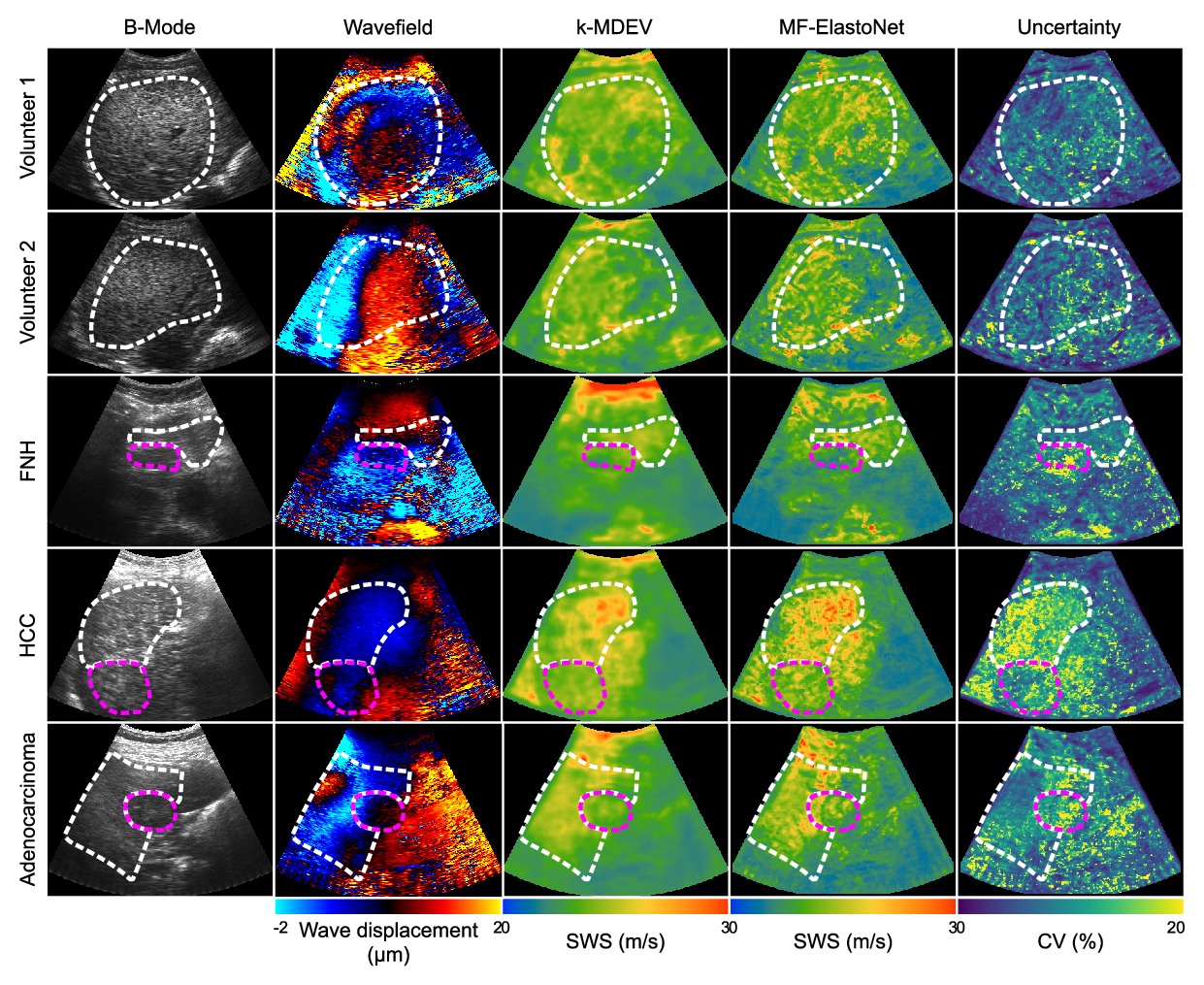}
\caption{B-mode, wavefield, SWS (shear wave speed) maps produced by k-MDEV and MF-ElastoNet, and CV (coefficient of variation) maps predicted by MF-ElastoNet are reported for two healthy volunteers, and three patients, including benign (23-year-old female with focal nodular hyperplasia) and malignant (81-year-old female with hepatocellular carcinoma and 49-year-old male with metastatic colorectal adenocarcinoma) cases. Liver parenchyma (white) and tumor (magenta) are additionally shown overlaid on all images.} \label{fig:liver_the}
\end{figure*}

\begin{figure*}
\centering
\includegraphics[width=0.8\textwidth]{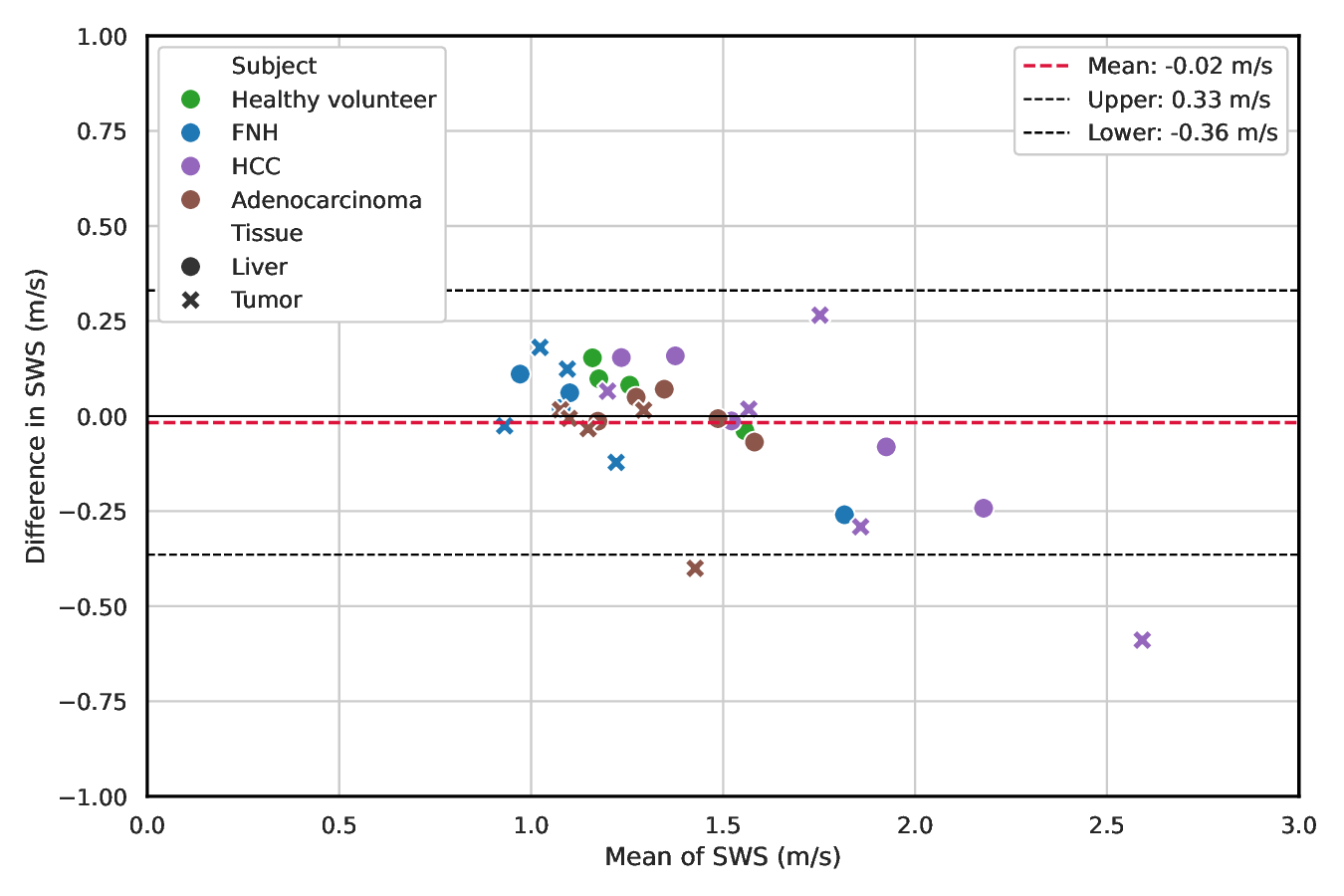}
\caption{Bland-Altman analysis of SWS (shear wave speed) values: MF-ElastoNet vs k-MDEV. Color denotes the subject groups (green: healthy volunteer, blue: focal nodular hyperplasia (FNH), purple: hepatocellular carcinoma (HCC), brown: adenocarcinoma), and marker shape tissue type (circle: liver parenchyma; cross: tumor).} \label{fig:results_liver_bland_altman}
\end{figure*}

The B-mode and wavefield images, predicted SWS maps by k-MDEV and MF-ElastoNet, as well as the predicted CV maps by MF-ElastoNet are displayed in Figure \ref{fig:liver_the}. The SWS maps produced by MF-ElastoNet appeared more detailed, while those produced by k-MDEV appeared smoother. k-MDEV and MF-ElastoNet showed excellent agreement in SWS values both in liver and tumor ROIs, as shown in Figure \ref{fig:results_liver_bland_altman}. Bland-Altman analysis yielded a mean bias of -0.02\,m/s (95\% limits of agreement: $-0.36$ to $0.33$\,m/s).

\section{Discussion and Conclusion}

In this work, we performed the first feasibility analysis of a multifrequency neural network-based wave inversion for ultrasound THE. We showed that MF-ElastoNet, a method that was first developed for MRE can be applied to ultrasound THE with minimal changes. We showed in the aMEM Challenge phantoms that MF-ElastoNet was able to resolve homogeneous areas as well as both soft and stiff inclusions. In \textit{in vivo} liver THE, healthy volunteers and patients with focal liver lesions showed excellent agreement in SWS values with k-MDEV used as our baseline, and even showed improved detail resolution, particularly relevant for tumor characterization \cite{Zhao_2026}.

These results were all the more promising considering that the model was applied in a pre-trained manner, with no fine-tuning and only adapting the preprocessing to ultrasound acquisitions. Downsampling was a necessary preprocessing step in this application of the pre-trained model. This step resulted in a loss of information, and adapting the model architecture and training data to THE should therefore be performed to fully leverage the detail resolution in this modality.

Other elastography modalities, such as optical multifrequency time-harmonic elastography \cite{Jordan_2024}, show similar challenges in that their resolution is even higher than in THE. Testing and adapting MF-ElastoNet to these challenging datasets could open the way for a unified cross-modality framework for multifrequency neural network-based wave inversion in elastography.

\begin{credits}

\subsubsection{\ackname} Financial support of the German Research Foundation (DFG, CRC1340 Matrix in Vision, GRK2260 BIOQIC, FOR5628) and the Medical Scientist Pilot Program of the Charité are gratefully acknowledged.

\subsubsection{\discintname} The authors have no competing interests to declare that are relevant to the content of this article.
\end{credits}
%
%
\bibliographystyle{splncs04}
\bibliography{reference}
\end{document}